%% file: main.tex
\RequirePackage[T1]{fontenc}
\documentclass[conference]{IEEEtran}
\IEEEoverridecommandlockouts
\usepackage{amsmath,amssymb,graphicx,booktabs,cite,url}
\usepackage[dvipsnames]{xcolor}
\usepackage{tikz}
\usepackage{pgfplots}
\pgfplotsset{compat=1.18}
\usepackage{placeins}
\usetikzlibrary{arrows.meta}
\newcommand{\E}{\mathcal{E}}
\newcommand{\D}{\mathcal{D}}
\newcommand{\cpp}{\mathrm{CPP}}
\newif\ifshowchanges
\showchangesfalse 
\ifshowchanges
\newcommand{\blue}[1]{{\textcolor[rgb]{0,0,1}{#1}}}
\else
\newcommand{\blue}[1]{#1}
\fi
\IEEEaftertitletext{\vspace{-1\baselineskip}}

\begin{document}
\bstctlcite{IEEEexample:BSTcontrol}
\title{\blue{Spatial Token Interleaving and Two-Pass\\BP Decoding for Semantic Image Transmission}\thanks{\blue{The code is available at \protect\url{https://github.com/Yifan-Usyd/STI-SemCom}.}}}
\author{\IEEEauthorblockN{Yifan Feng, Chentao Yue, Haiyao Yu, Gaoyang Pang, Xinyu Chen, Hao Chang, Branka Vucetic, and Yonghui Li}
\IEEEauthorblockA{School of Electrical and Computer Engineering, The University of Sydney, Australia}}
\maketitle
\begin{abstract}
\blue{Semantic error control coding (SECC) can correct the residual errors of a standard channel code with a foundation model that supplies a prior on the source content, but existing SECC schemes are designed for text. In this paper, we develop SECC for image transmission with spatial token interleaving and a two-pass belief propagation (BP) schedule. At the transmitter, a vector-quantized (VQ) tokenizer converts an image into tokens, which are transmitted in multiple low-density parity-check (LDPC) code blocks of fifth-generation New Radio (5G NR). In raster order, a block that BP fails to decode corrupts consecutive token rows, so the interleaver reorders the tokens such that adjacent tokens are carried by different blocks, without additional transmitted bits. At the receiver, after the first BP pass, token recovery fuses the token probabilities given by BP with the predictions of the autoregressive model Emu3-Gen, and the second BP pass decodes each block again with the fused distributions as priors on the information bits. On Kodak24 at 0.5 channel uses per pixel and a signal-to-noise ratio (SNR) of 0.15~dB, the second pass removes 92\% of the token errors left by token recovery, and interleaving reduces the final token error rate by 88\%. At medium to high SNR, the proposed receiver exceeds deep joint source--channel coding baselines with the same channel uses by at least 1.2~dB in peak signal-to-noise ratio.}
\end{abstract}

\vspace{-0.5em}
\section{Introduction}\label{sec:intro}
\vspace{-0.5em}

\blue{Token communications~\cite{qiao2025tokcom} represent an image by discrete tokens and transmit them over a wireless link~\cite{fu2023vqdeepsc}. A vector-quantized (VQ) tokenizer~\cite{vandenoord2017vqvae} divides the image into a grid of patches and represents each patch by the index of the nearest entry of a learned codebook. Unlike deep joint source--channel coding (JSCC)~\cite{bourtsoulatze2019deepjscc,yang2023swinjscc}, which maps an image directly to channel symbols, token communications can protect the tokens with the channel codes of existing standards~\cite{liu2025textguided}. The data channels of fifth-generation New Radio (5G NR) use low-density parity-check (LDPC) codes and segment a transport block longer than 8448 bits into code blocks, which are encoded and decoded separately~\cite{3gpp38212}. An image usually needs more bits than one code block. With the Emu3 tokenizer~\cite{wang2024emu3}, which represents each $8\times8$ patch by one index, any image larger than about $200\times200$ pixels needs more than 8448 bits.}

\blue{Belief propagation (BP) decodes the code blocks separately, and near the BP threshold, it leaves residual errors in some blocks while decoding the others correctly. Semantic error control coding (SECC)~\cite{yue2026secc} can correct such errors with a foundation model that supplies a prior on the source content, i.e., the probability of each source symbol given its context, while the channel code keeps its algebraic structure. Hao~et~al.~\cite{hao2025shortwins} showed for text that multiple short codes with semantic correction outperform a single long code, because a decoding failure is confined to one block and the other blocks provide the context for its correction. For images, DiffCom~\cite{wang2025diffcom} guides diffusion posterior sampling with the received signal, and text-guided token prediction reconstructs images whose tokens are protected by 5G NR polar codes~\cite{liu2025textguided}. Autoregressive (AR) models such as Emu3~\cite{wang2024emu3} provide the prior for image tokens as the probability of each codebook index given the preceding tokens.}

\blue{The SECC schemes in~\cite{li2026llmviterbi,wang2026clsec,hao2026iterative,hao2025shortwins} are designed for text. For images, two problems remain, which concern the context of an erroneous token and the decoding of the model output. An image prior corrects an erroneous token from the tokens around it on the grid. In raster order, i.e., row by row, a code block carries consecutive token rows, e.g., about nine for a $512\times512$ image with the Emu3 tokenizer. A failed block then corrupts a band of the image, and the correctly decoded blocks provide context only for the token rows at the two edges of the band. Bit-interleaved coded modulation~\cite{caire1998bicm} permutes coded bits rather than tokens, and character interleaving for text~\cite{hao2026iterative} separates neighbors along one dimension only.} \blue{Moreover, the model output is not decoded again in~\cite{li2026llmviterbi,wang2026clsec,liu2025textguided}, where the model selects among the candidate paths of a list Viterbi decoder~\cite{li2026llmviterbi} or corrects the decoded symbols~\cite{wang2026clsec,liu2025textguided}, so the parity constraints of the code cannot correct the errors that the model leaves.}

\blue{In this paper, we propose spatial token interleaving and a two-pass BP schedule for semantic image transmission. At the transmitter, the Emu3 tokenizer converts an image into tokens. The interleaver reorders the tokens by a fixed permutation known to the receiver, without additional transmitted bits, such that adjacent tokens are carried by different blocks and thus have independent errors. The interleaved tokens are then transmitted in multiple 5G NR LDPC code blocks. At the receiver, after the first pass of BP decoding, token recovery fuses the token probabilities given by BP with the predictions of the Emu3-Gen model~\cite{wang2024emu3}. The second BP pass decodes each block again with the fused distributions as prior log-likelihood ratios (LLRs) of the information bits, so that the parity constraints correct the errors left by token recovery. On Kodak24 at 0.5 channel uses per pixel (CPP) and a signal-to-noise ratio (SNR) of 0.15~dB, the second pass removes 92\% of these errors, and interleaving reduces the final token error rate (TER) by 88\%. At medium to high SNR, the peak signal-to-noise ratio (PSNR) of the proposed receiver exceeds those of SwinJSCC~\cite{yang2023swinjscc} and FAJSCC~\cite{choi2025fajscc} by at least 1.2~dB at the same CPP.}

\vspace{-0.5em}
\section{System Model}\label{sec:system}
\vspace{-0.5em}
\begin{figure}[!t]
\centering
\input{figures/tikz/fig_system_2}%
\caption{System overview. }
\label{fig:system}
\vspace{-1em}
\end{figure}
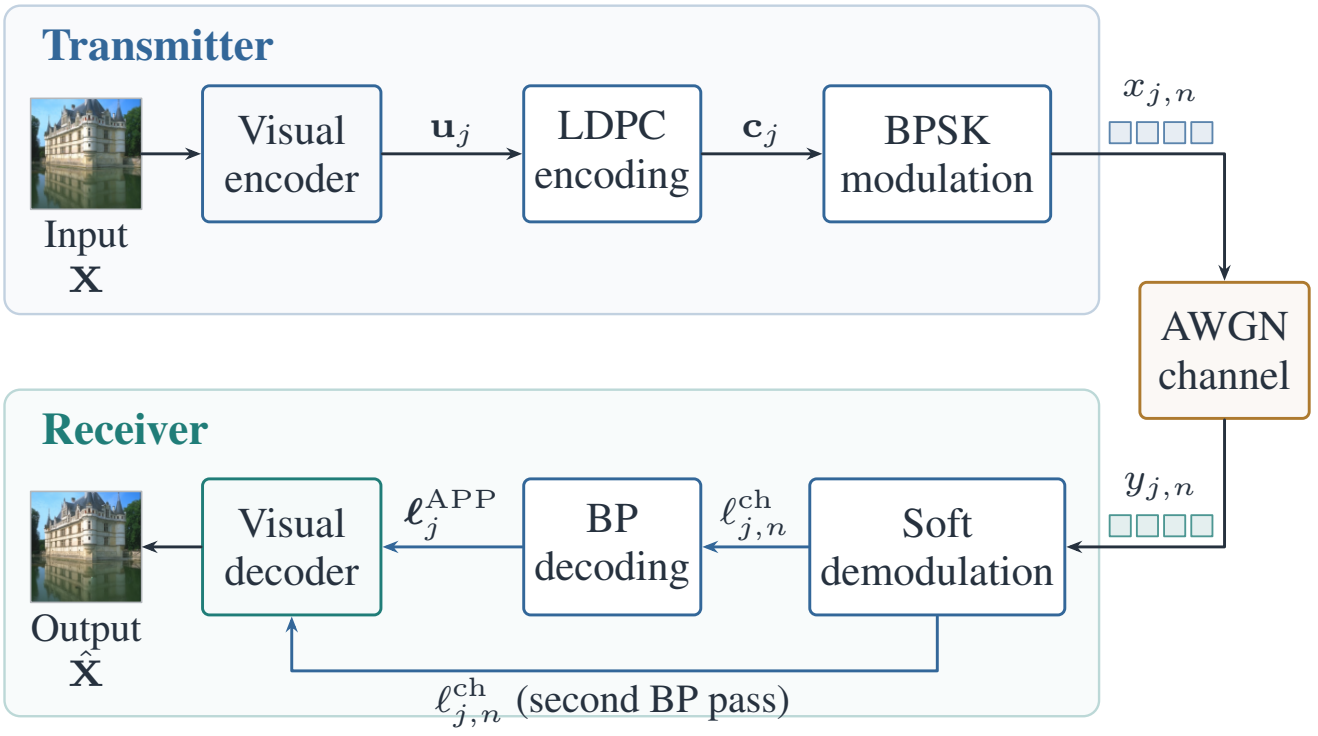

\subsection{\protect\blue{Encoding and Channel}}

As illustrated in Fig.~\ref{fig:system}, we consider point-to-point image transmission over an additive white Gaussian noise (AWGN) channel. At the transmitter, the visual encoder $\E$ uses a tokenizer to map a red--green--blue (RGB) image $\mathbf{X}\in\mathbb{R}^{H\times W\times3}$ to a sequence of $T$ indices in a fixed codebook $\mathcal C$. These indices are interleaved, mapped to binary labels \blue{of $M$ bits} and partitioned into $B$ consecutive information blocks, each containing $K$ bits. The process can be expressed as
\begin{equation}
[\mathbf u_1,\ldots,\mathbf u_B]
= \E (\mathbf X),
\end{equation}
where $\mathbf u_j\in\{0,1\}^{1\times K}$ is the $j$-th information block, for $j=1,\ldots,B$.
These information blocks are then LDPC-encoded by
\begin{equation}
[\mathbf c_1,\ldots,\mathbf c_B]
=[\mathbf u_1\mathbf G,\ldots,\mathbf u_B\mathbf G]
\in\{0,1\}^{1\times BN},
\end{equation}
where $\mathbf G\in\{0,1\}^{K\times N}$ is the generator matrix, with multiplication over $\mathrm{GF}(2)$, and $\mathbf{c}_j=[c_{j,1},\ldots,c_{j,N}]$ is the \blue{$j$-th codeword}. Hence, the code rate is $R=K/N$. \blue{As for the 5G NR LDPC codes~\cite{3gpp38212}, $\mathbf G$ is systematic, i.e., the first $K$ coded bits of $\mathbf c_j$ are the information bits $\mathbf u_j$.}
Binary phase-shift keying (BPSK) maps each coded bit $c_{j,n}$ to the transmitted symbol
$x_{j,n}=1-2c_{j,n}$, for $n=1,\ldots,N$. \blue{An image thus occupies $BN$ real channel dimensions, i.e., $\cpp=BN/(HW)$.}

At the receiver, the received symbol \blue{is $y_{j,n}=x_{j,n}+w_{j,n}$,}
where $w_{j,n}\sim\mathcal N(0,\blue{\sigma^2})$ is \blue{independent Gaussian noise}. With unit symbol energy\blue{\ $E_{\mathrm s}=1$}, the \blue{SNR per real channel dimension is $\gamma=1/\sigma^2=2E_{\mathrm s}/N_0$, where $N_0=2\sigma^2$ is the noise power spectral density}.
The received symbols are then soft-demodulated to obtain channel \blue{LLRs} for the coded bits. Specifically, with a bit-1 over bit-0 LLR convention, the channel LLR for the $n$th received symbol in block $j$ is
$\ell^{\mathrm{ch}}_{j,n}=\blue{-2y_{j,n}/\sigma^2}$.
\vspace{-0.3em}
\subsection{\protect\blue{BP Decoding}}\label{ssec:bp}
\vspace{-0.3em}
\blue{The receiver decodes each block separately with BP~\cite{richardson2008modern}, and we omit the block index $j$ in \eqref{eq:bp-check}--\eqref{eq:app}. For the parity checks $r=1,\ldots,N-K$ of the code, let $\partial r$ be the set of coded bits in check $r$ and $\partial n$ the set of checks on coded bit $n$. BP takes input LLRs $\ell^{\mathrm{in}}_n$ of the $N$ coded bits, which are the channel LLRs in the first pass, i.e., $\ell^{\mathrm{in}}_n=\ell^{\mathrm{ch}}_n$. It initializes $\mu_{n\to r}=\ell^{\mathrm{in}}_n$ for $r\in\partial n$ and, in each of a fixed number of iterations, computes the messages}
\blue{\begin{align}
\mu_{r\to n}&=-2\,\mathrm{artanh}\prod_{n'\in\partial r\setminus\{n\}}\tanh\Bigl(-\frac{\mu_{n'\to r}}{2}\Bigr),\label{eq:bp-check}\\
\mu_{n\to r}&=\ell^{\mathrm{in}}_n+\sum_{r'\in\partial n\setminus\{r\}}\mu_{r'\to n},\label{eq:bp-var}
\end{align}}%
\blue{where the signs in \eqref{eq:bp-check} follow from the bit-1 over bit-0 convention. After the last iteration, BP outputs the approximate a posteriori probability (APP) LLRs}
\blue{\begin{equation}
\ell^{\mathrm{APP}}_n=\ell^{\mathrm{in}}_n+\sum_{r\in\partial n}\mu_{r\to n},\quad n=1,\ldots,N,
\label{eq:app}
\end{equation}}%
\blue{and the hard decisions $\hat c_n=\mathbf 1\{\ell^{\mathrm{APP}}_n>0\}$.} \blue{The APP LLRs of the information bits $n=1,\ldots,K$ of block $j$ in the first pass form} $\boldsymbol\ell_j^{\mathrm{APP}}\in\mathbb R^{1\times K}$.
The APP LLRs from all $B$ blocks are collected in $\boldsymbol\ell^{\mathrm{APP}}=[\boldsymbol\ell_j^{\mathrm{APP}}]_{j=1}^B\in\mathbb R^{B\times K}$ and processed by the visual decoder $\D$\blue{, which also uses the channel LLRs $\boldsymbol\ell^{\mathrm{ch}}=[\ell^{\mathrm{ch}}_{j,n}]\in\mathbb R^{B\times N}$ in a second BP pass (Section~\ref{ssec:second-bp}),} to reconstruct the image as $\hat{\mathbf X}=\D(\boldsymbol\ell^{\mathrm{APP}}\blue{,\boldsymbol\ell^{\mathrm{ch}}})$.

\blue{Let $\delta_j$ and $\rho_j$ be the fractions of erroneous bits and of erroneous tokens among the labels that lie in block $j$ after the first pass. We say that BP fails on block $j$ if $\delta_j>0$. Since the binary-input AWGN channel and BP are symmetric, the distribution of the error pattern after BP does not depend on the transmitted codeword~\cite{richardson2008modern}. As BP decodes block $j$ from its own received symbols and the noise is independent across blocks, the error patterns of the $B$ blocks, and hence $\delta_1,\ldots,\delta_B$, are independent, and the erroneous bits of an image lie in its failed blocks. A token is erroneous if and only if at least one bit of its label is in error. For a label of $M$ bits with $e$ erroneous bits, $e/M\le\mathbf 1\{e\ge1\}\le e$, and averaging over the labels in block $j$ gives}
\blue{\begin{equation}
\delta_j\le\rho_j\le\min\{1,M\delta_j\}.
\label{eq:token-bit}
\end{equation}}%
\blue{In the waterfall region of the code, even at an SNR at which BP decodes most blocks correctly, its failures are large errors~\cite{richardson2008modern}, i.e., the fraction $\delta_j$ of a failed block does not vanish as $K$ grows. By \eqref{eq:token-bit}, a failed block then corrupts a fraction of at least $\delta_j$ of its tokens. Without interleaving, these tokens form consecutive token rows, i.e., a band of the image, and an erroneous token inside the band has no neighbor in a correctly decoded block, leaving no reliable context for its recovery. The token interleaver of Section~\ref{ssec:permutation} assigns adjacent tokens to different blocks.}

\vspace{-0.5em}
\section{Visual Encoder Design}\label{sec:method}
\vspace{-0.3em}

\begin{figure}[!t]
\centering
\resizebox{\columnwidth}{!}{\input{figures/tikz/fig_encoder_2}}
\caption{Visual encoder \blue{and the LDPC encoding of each block}.}
\vspace{-1em}
\label{fig:encoder-design}
\end{figure}
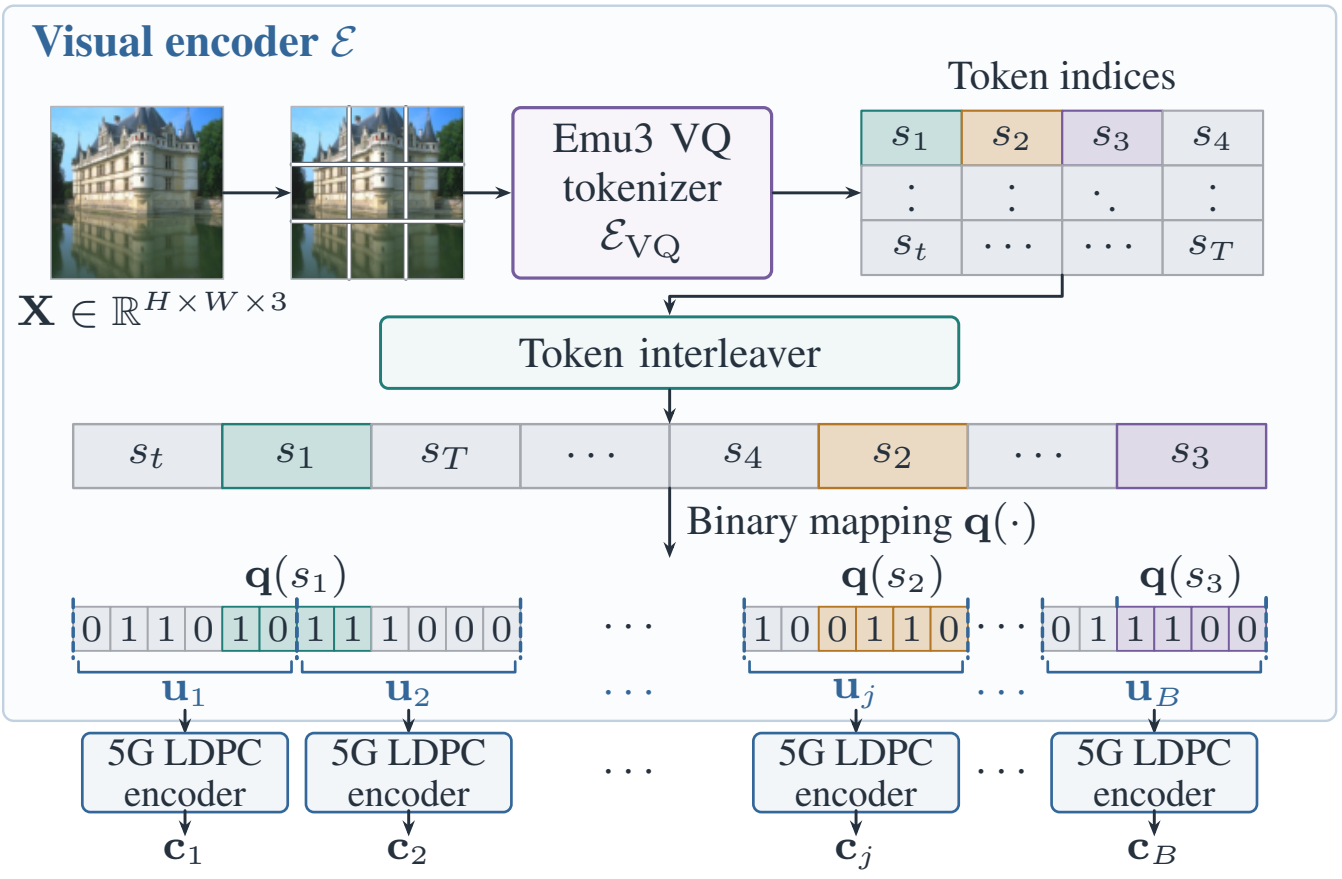

As illustrated in Fig.~\ref{fig:encoder-design}, the visual encoder $\E$ converts an image into $B$ information blocks through Emu3 VQ tokenization, interleaving, and binary mapping.
\vspace{-0.3em}
\subsection{Emu3 VQ Tokenization}
\vspace{-0.3em}

For an image $\mathbf{X}\in\mathbb{R}^{H\times W\times 3}$, the pretrained Emu3 VQ tokenizer $\E_{\mathrm{VQ}}$~\cite{wang2024emu3} extracts a spatial latent grid with $\blue{H'}$ rows and $\blue{W'}$ columns, giving $T=\blue{H'W'}$ token positions.

\blue{At} each position $t\in \{1,2,\ldots, T\}$, vector quantization selects the nearest embedding \blue{of the Emu3 codebook} and returns its index $s_t\in\{0,\ldots,V-1\}$\blue{. For images, the Emu3 tokenizer uses only the first $2^{14}$ of its $2^{15}$ codebook entries~\cite{emu3issue59}, and we take these entries as the codebook $\mathcal C$, so that $V=|\mathcal C|=2^{14}$ and each index has a 14-bit label}. Collecting these indices in raster order yields the token sequence $\mathbf s=[s_1,\ldots,s_T]$, expressed as
\begin{equation}
\mathbf s=\E_{\mathrm{VQ}}(\mathbf X)
\in\{0,\ldots,V-1\}^{1\times T}.
\end{equation}

\vspace{-0.3em}
\subsection{Interleaving and Binary Mapping}
\vspace{-0.3em}

We apply token-level interleaving to distribute neighboring positions across different information blocks\blue{. The neighbors of an erroneous token then lie in other blocks, which are mostly decoded correctly, and give the autoregressive model at the receiver context to recover the token despite transmission errors (Section~\ref{ssec:bp})}. Let $\pi$ be a fixed permutation of $\{1,\ldots,T\}$, independent of image content and shared by the transmitter and receiver. The interleaver produces \blue{$\mathbf s_{\pi}=[s_{\pi(1)},\ldots,s_{\pi(T)}]$,} where $\pi(i)$ denotes the original spatial position of the token carried in transmission slot $i$.

Binary mapping then converts each index $s_{\pi(i)}\in\{0,\ldots,V-1\}$ into \blue{its} $M$-bit label $\mathbf q(s_{\pi(i)})\in\{0,1\}^{1\times M}$\blue{, where $M=\log_2 V$}. Concatenating the labels in transmission order yields a $TM$-bit information stream. We partition this stream into $B$ consecutive blocks $\mathbf u_j\in\{0,1\}^{1\times K}$, for $j=1,\ldots,B$, with $BK=TM$, \blue{so that a label can straddle two blocks, and obtain}
\begin{equation}
\relax[\mathbf u_1,\ldots,\mathbf u_B]
=[\mathbf q(s_{\pi(1)}),\ldots,\mathbf q(s_{\pi(T)})]\in\{0,1\}^{1\times BK}.
\label{eq:encoder-output}
\end{equation}

\vspace{-0.3em}
\subsection{\protect\blue{Interleaver Design}}\label{ssec:permutation}
\vspace{-0.3em}
\blue{By \eqref{eq:encoder-output}, the label in slot $i$ occupies bits $(i-1)M+1$ to $iM$ of the information stream, and its bits lie in the blocks}
\blue{\begin{equation}
\mathcal J(i)=\bigl\{j:\lceil((i-1)M+1)/K\rceil\le j\le\lceil iM/K\rceil\bigr\},
\label{eq:label-blocks}
\end{equation}}%
\blue{i.e., in one block, or in two if the label straddles a block boundary. Token $t$ is carried in slot $\pi^{-1}(t)$, and we write $\mathcal J_\pi(t)=\mathcal J(\pi^{-1}(t))$. Let $\mathcal A_t$ be the event that the label of token $t$ has an erroneous bit after the first pass. Since the error patterns of different blocks are independent (Section~\ref{ssec:bp}), for any two tokens $t$ and $t'$,}
\blue{\begin{equation}
\mathcal J_\pi(t)\cap\mathcal J_\pi(t')=\emptyset
\;\Longrightarrow\;
\Pr(\mathcal A_{t'}\mid\mathcal A_t)=\Pr(\mathcal A_{t'}).
\label{eq:indep-tokens}
\end{equation}}%
\blue{Let $\mathcal P$ be the set of the $H'(W'-1)+(H'-1)W'$ pairs of horizontally or vertically adjacent tokens. We require}
\blue{\begin{equation}
\kappa(\pi)=\sum_{\{t,t'\}\in\mathcal P}\mathbf 1\{\mathcal J_\pi(t)\cap\mathcal J_\pi(t')\neq\emptyset\}=0,
\label{eq:shared-pairs}
\end{equation}}%
\blue{i.e., no two adjacent tokens have bits in a common block, so that by \eqref{eq:indep-tokens}, the errors of any two adjacent tokens are independent. To also spread the labels of larger neighborhoods evenly over the blocks, let $\Omega_k$ be the set of the $(H'-k+1)(W'-k+1)$ windows of $k\times k$ tokens, let $h_j(\omega)$ be the number of tokens $t$ in window $\omega$ with $j\in\mathcal J_\pi(t)$, and let $\bar h(\omega)$ be the mean of $h_1(\omega),\ldots,h_B(\omega)$. We measure the spread by the normalized variance}
\blue{\begin{equation}
\Psi_k(\pi)=\frac{1}{|\Omega_k|}\sum_{\omega\in\Omega_k}\frac{1}{Bk^2}\sum_{j=1}^{B}\bigl(h_j(\omega)-\bar h(\omega)\bigr)^2,
\label{eq:window-variance}
\end{equation}}%
\blue{which is small if the labels of each window are spread evenly over the blocks.}

\blue{We construct $\pi$ by a local search over the grid positions and the block structure. Starting from the affine permutation $\pi_0(i)=1+\bigl(33(i-1)\bmod T\bigr)$, which satisfies $\kappa(\pi_0)=0$, the search proposes to exchange the slots of two randomly chosen tokens and accepts the exchange only if it decreases $\bigl(\kappa(\pi),\Psi_3(\pi),\Psi_5(\pi),\Psi_9(\pi)\bigr)$ in lexicographic order. With a fixed random seed, 807 of 6,000 proposed exchanges are accepted. The resulting $\pi$ has $\kappa(\pi)=0$, and $\Psi_3$ decreases from 0.147 for $\pi_0$ to 0.078, $\Psi_5$ from 0.247 to 0.077, and $\Psi_9$ from 0.276 to 0.071.}

\vspace{-0.5em}
\section{Visual Decoder Design}\label{sec:decoder}
\vspace{-0.3em}
\begin{figure}[!t]
\centering
\resizebox{\columnwidth}{!}{\input{figures/tikz/fig_decoder_2}}
\caption{\blue{First BP pass and visual} decoder \blue{with the second BP pass.}}
\vspace{-1em}
\label{fig:decoder-design}
\end{figure}
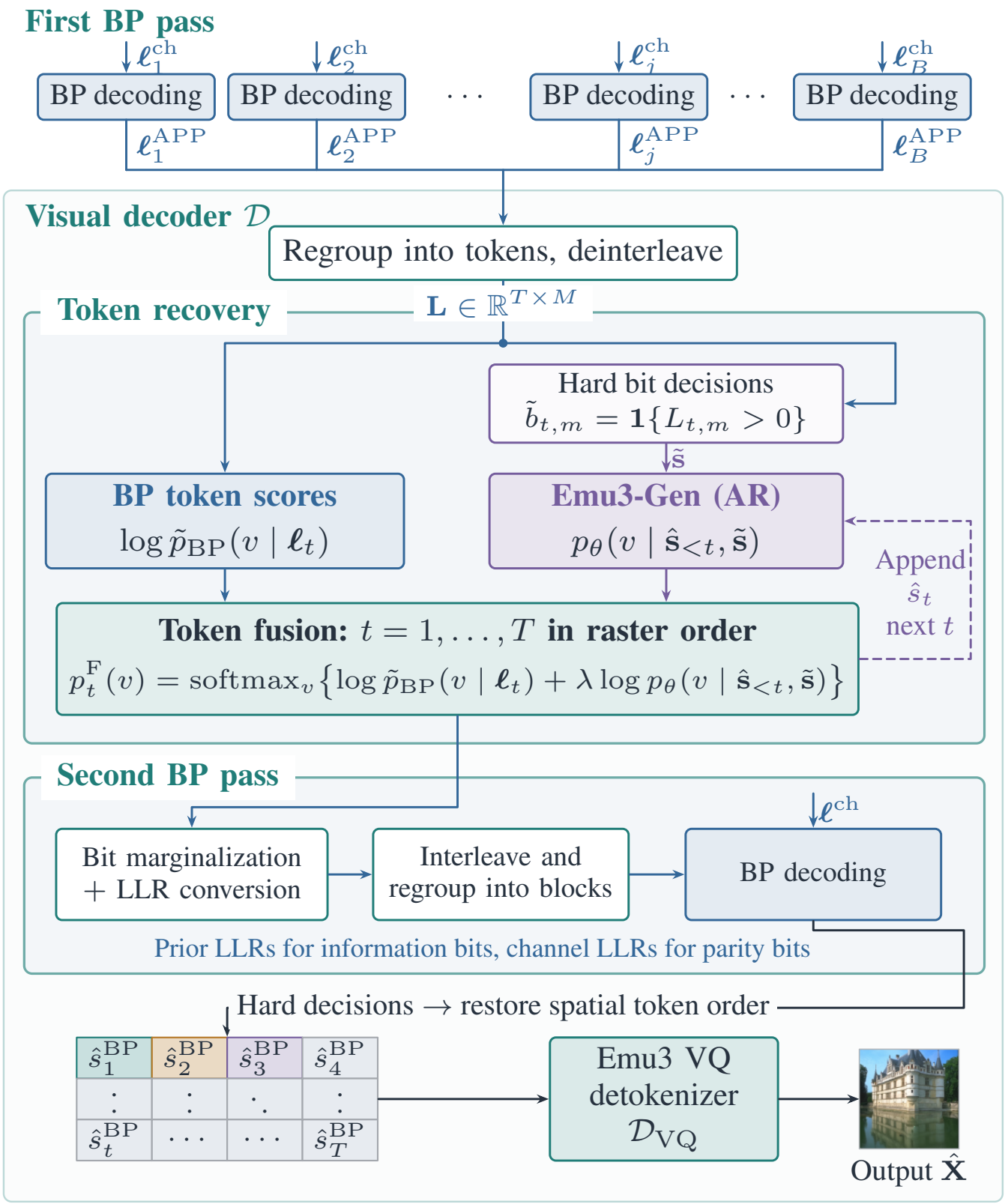

As shown in Fig.~\ref{fig:decoder-design}, \blue{the receiver first decodes each block in the first BP pass. The visual decoder $\D$ then restores the spatial token order of the resulting APP LLRs (Section~\ref{ssec:deinterleave}), recovers the token indices from these LLRs and the predictions of Emu3-Gen (Section~\ref{ssec:token-recovery}), decodes each block again in the second BP pass with the fused token distributions as priors (Section~\ref{ssec:second-bp}), and maps the decoded tokens to pixels (Section~\ref{ssec:reconstruction}).}

\vspace{-0.3em}
\subsection{\protect\blue{First BP Pass and} Deinterleaving}\label{ssec:deinterleave}
\vspace{-0.3em}

\blue{The first BP pass applies \eqref{eq:bp-check}--\eqref{eq:app} to each block $j$ with $\ell^{\mathrm{in}}_n=\ell^{\mathrm{ch}}_{j,n}$ and yields $\boldsymbol\ell^{\mathrm{APP}}_1,\ldots,\boldsymbol\ell^{\mathrm{APP}}_B$.} Following the grouping relation in \eqref{eq:encoder-output}, we concatenate the $B$ information-bit APP LLR vectors and regroup the resulting sequence into $T$ token-wise vectors
\begin{equation}
[\mathbf a_1,\ldots,\mathbf a_T]=[\boldsymbol\ell_1^{\mathrm{APP}},\ldots,\boldsymbol\ell_B^{\mathrm{APP}}]\in\mathbb R^{1\times TM},
\label{eq:regroup}
\end{equation}
where $\mathbf a_i=[a_{i,1},\ldots,a_{i,M}]\in\mathbb R^{1\times M}$, $\forall i\in\{1,\ldots,T\}$, contains the APP LLRs for the $M$ bits of the token transmitted in \blue{slot} $i$.

The token deinterleaver then restores the original spatial order. Since transmission slot $i$ carries the token from spatial position $t=\pi(i)$, the slot associated with position $t$ is $i=\pi^{-1}(t)$. The deinterleaver therefore produces
\begin{equation}
\boldsymbol\ell_t=\mathbf a_{\pi^{-1}(t)}\in\mathbb{R}^{1\times M}.
\label{eq:inverse}
\end{equation}
After deinterleaving all $T$ token-wise vectors, we form the spatially aligned LLR matrix as $\mathbf L=[\boldsymbol\ell_1^{\mathsf T},\ldots,
\boldsymbol\ell_T^{\mathsf T}]^{\mathsf T}\in\mathbb R^{T\times M}$, where \blue{its $(t,m)$th entry $L_{t,m}$} is the LLR for bit $m$ of the token at spatial position $t$.

\vspace{-0.3em}
\subsection{Token Recovery}\label{ssec:token-recovery}
\vspace{-0.3em}

Given $\mathbf L$, \blue{token recovery combines BP token scores with Emu3-Gen predictions to recover the token indices in raster order}. The recovered sequence is denoted by $\hat{\mathbf s}=[\hat s_1,\ldots,\hat s_T]$.
\subsubsection{BP token scoring}
\blue{At position $t$, the BP branch scores each candidate index $v\in\{0,\ldots,V-1\}$ by the probability of its label $\mathbf q(v)$ under the LLRs $\boldsymbol\ell_t$. Since BP provides only bit marginals, the $M$ bits are treated as independent, i.e.,}
\begin{equation}
\tilde p_{\mathrm{BP}}(v\mid\boldsymbol\ell_t)
=\prod_{m=1}^{M}
p_{t,m}^{\blue{q_m(v)}}
(1-p_{t,m})^{1-\blue{q_m(v)}},\label{eq:bp_prob}
\end{equation}
where $\blue{q_m(v)}\in\{0,1\}$ is the $m$th bit of $\mathbf q(v)$ \blue{and $p_{t,m}=(1+e^{-L_{t,m}})^{-1}$ is the probability that bit $m$ of the token at position $t$ equals one.}

\subsubsection{Emu3 token scoring}
The Emu3 token scoring branch \blue{scores the candidates from visual context. Its input is the hard token sequence, obtained from the hard bit decisions} $\tilde b_{t,m}=\mathbf 1\{L_{t,m}>0\}$ \blue{as}
\begin{equation}
\tilde s_{t}
=\mathbf q^{-1}
([\tilde b_{t,1},\ldots,\tilde b_{t,M}]),
\end{equation}
where $\mathbf q^{-1}$ maps a binary label to its codebook index. Collecting these indices yields the hard token sequence $\tilde{\mathbf s}=[\tilde s_1,\ldots,\tilde s_T]$.

\blue{We use} the Emu3-Gen causal Transformer~\cite{wang2024emu3}\blue{, which generates an image token by token in raster order over a vocabulary that includes the $V$ codebook indices}.
\blue{As shown in Fig.~\ref{fig:decoder-design}, its input consists of} the complete hard token sequence $\tilde{\mathbf s}$ \blue{followed by} the recovered prefix $\hat{\mathbf s}_{<t}=[\hat s_1,\ldots,\hat s_{t-1}]$\blue{, with image-boundary and row-separator tokens marking the grid structure}. \blue{Since $\tilde{\mathbf s}$ precedes the prefix, the prediction at position $t$ attends to the observations of the whole image, including those to the right of and below position $t$. With interleaving, the four neighbors of position $t$ are carried by blocks other than its own (Section~\ref{ssec:permutation}).} At each position, a softmax over the $V$ visual-token logits defines $p_\theta(v\mid\hat{\mathbf s}_{<t},\tilde{\mathbf s})$, the probability assigned to codebook index $v$. \blue{Both branches thus score the same $V$ codebook entries, and the LLR magnitudes enter only through the BP branch.}

\subsubsection{Score fusion \protect\blue{and} token selection}
\blue{At position $t$, the fused distribution of the two branches is}
\begin{equation}
\blue{p^{\mathrm F}_t(v)\propto\tilde p_{\mathrm{BP}}(v\mid\boldsymbol\ell_t)\,
p_\theta(v\mid\hat{\mathbf s}_{<t},\tilde{\mathbf s})^{\lambda},}
\label{eq:fused}
\end{equation}
\blue{normalized over the $V$ candidates, and the recovered index is}
\begin{equation}
\hat s_t=\arg\max_{v\in\{0,\ldots,V-1\}}\bigl\{\log\tilde p_{\mathrm{BP}}(v\mid\boldsymbol\ell_t)
+\lambda\log p_\theta(v\mid\hat{\mathbf s}_{<t},\tilde{\mathbf s})\bigr\},
\label{eq:score-fusion-recovery}
\end{equation}
where $\lambda\ge0$ is an empirically selected weight of the Emu3 prediction.

After selecting $\hat s_t$, we append it to the recovered prefix, i.e., $\hat{\mathbf s}_{<t+1}=[\hat{\mathbf s}_{<t},\hat s_t]$, which is used by Emu3-Gen to predict the next token. Starting with an empty prefix at $t=1$, this process follows raster order until all $T$ tokens are recovered, yielding $\hat{\mathbf s}=[\hat s_1,\ldots,\hat s_T]$. The hard token sequence $\tilde{\mathbf s}$ and \blue{the LLRs $\mathbf L$} remain fixed throughout recovery.

\vspace{-0.3em}
\subsection{\protect\blue{Second BP Pass}}\label{ssec:second-bp}
\vspace{-0.3em}

\blue{Token recovery corrects erroneous tokens from their context but does not use the parity constraints of the code. The second BP pass returns the fused distributions to the LDPC decoder as priors. For bit $m$ of the token at position $t$, marginalizing \eqref{eq:fused} over the codebook gives the prior LLR}
\begin{equation}
\blue{L^{\mathrm F}_{t,m}=\ln\frac{\sum_{v:\,q_m(v)=1}p^{\mathrm F}_t(v)}{\sum_{v:\,q_m(v)=0}p^{\mathrm F}_t(v)}}
\label{eq:prior-llr}
\end{equation}
\blue{in the bit-1 over bit-0 convention of Section~\ref{sec:system}. Interleaving these LLRs with $\pi$ and grouping them into blocks, i.e., inverting \eqref{eq:inverse} and \eqref{eq:regroup}, yields the prior LLRs $\boldsymbol\ell^{\mathrm F}_j=[\ell^{\mathrm F}_{j,1},\ldots,\ell^{\mathrm F}_{j,K}]$ of the information bits of block $j$. The second pass decodes each block $j$ anew by \eqref{eq:bp-check}--\eqref{eq:app} with the input LLRs}
\blue{\begin{equation}
\ell^{\mathrm{in}}_n=\begin{cases}
\ell^{\mathrm F}_{j,n}, & n=1,\ldots,K,\\
\ell^{\mathrm{ch}}_{j,n}, & n=K+1,\ldots,N.
\end{cases}
\label{eq:second-pass-input}
\end{equation}}%
\blue{The prior LLRs replace the channel LLRs of the information bits instead of being added to them, because $p^{\mathrm F}_t$ is computed from the APP LLRs in \eqref{eq:bp_prob}, which already contain these channel LLRs. The hard decisions of the second pass are regrouped by \eqref{eq:regroup}, deinterleaved by \eqref{eq:inverse} and mapped to indices by $\mathbf q^{-1}$, which yields the decoded token sequence $\hat{\mathbf s}^{\mathrm{BP}}=[\hat s^{\mathrm{BP}}_1,\ldots,\hat s^{\mathrm{BP}}_T]$. The two passes form the two-pass BP schedule, which requires no additional transmission.}

\vspace{-0.3em}
\subsection{Image Reconstruction}\label{ssec:reconstruction}
\vspace{-0.3em}

\blue{We} reshape the sequence $\blue{\hat{\mathbf s}^{\mathrm{BP}}}$ row by row into an $\blue{H'\times W'}$ token grid. The grid is then passed to the fine-tuned Emu3 VQ detokenizer $\mathcal D_{\mathrm{VQ}}$. Within the detokenizer, each index selects an embedding from the fixed codebook $\mathcal C$, and the resulting spatial features are processed by the image decoding network\blue{, which outputs}
\begin{equation}
\hat{\mathbf X}=\D_{\mathrm{VQ}}(\blue{\hat{\mathbf s}^{\mathrm{BP}}})
\in\mathbb R^{H\times W\times3}.
\label{eq:image-reconstruction}
\end{equation}

\vspace{-0.3em}
\subsection{Training of the Visual Decoder}\label{ssec:decoder-training}
\vspace{-0.3em}

The visual decoder contains two neural models\blue{, namely} Emu3-Gen and the VQ detokenizer. We fine-tune \blue{them} separately. The pretrained VQ tokenizer $\E_{\mathrm{VQ}}$ and the codebook $\mathcal C$ remain fixed.

\subsubsection{Training of the Emu3-Gen model}
We adapt Emu3-Gen using low-rank adaptation (LoRA)~\cite{hu2022lora}, which updates the inserted \blue{low-rank} parameters $\theta$ while keeping the backbone weights frozen. \blue{The hard token sequences for training are obtained by passing the training images through the encoder of Section~\ref{sec:method}, the coded channel and the first BP pass.}

During fine-tuning, we use the hard token sequence $\tilde{\mathbf s}$ and the \blue{clean} prefix $\mathbf s_{<t}=[s_1,\ldots,s_{t-1}]$ to predict the clean token $s_t$ at each position $t$. The training loss is the per-token cross-entropy (CE)
\begin{equation}
\mathcal L_{\mathrm{CE}}=-\frac{1}{T}\sum_{t=1}^{T}\log P_\theta(\nu(s_t)\mid\mathbf s_{<t},\tilde{\mathbf s})\blue{,}
\end{equation}
\blue{where} $\nu(s_t)$ is the vocabulary identifier \blue{of} codebook index $s_t$, and $P_\theta$ is normalized over the full text--visual vocabulary \blue{of Emu3-Gen, whereas} $p_\theta$ \blue{at inference} is normalized over \blue{the $V$} visual identifiers. Training uses clean prefixes\blue{, whereas} inference uses \blue{the recovered prefix}.

\subsubsection{Training of the VQ detokenizer}
Even with correct token indices, image quality depends on how the fixed codebook representation is mapped to pixels. \blue{We therefore} fine-tune the parameters $\phi$ of the image decoding network and the post-quantization convolution within the detokenizer using \blue{clean} token--image pairs $(\mathbf s,\mathbf X)$, where $\mathbf s=\mathcal E_{\mathrm{VQ}}(\mathbf X)$\blue{, with the loss}
\begin{equation}
\mathcal L_{\mathrm{R}}=\mathrm{MSE}\bigl(\D_{\mathrm{VQ}}(\mathbf s),\mathbf X\bigr)+\mathrm{LPIPS}\bigl(\D_{\mathrm{VQ}}(\mathbf s),\mathbf X\bigr),
\label{eq:detokenizer-training}
\end{equation}
where \blue{MSE and LPIPS denote the mean squared error and the learned perceptual image patch similarity~\cite{zhang2018lpips}}.

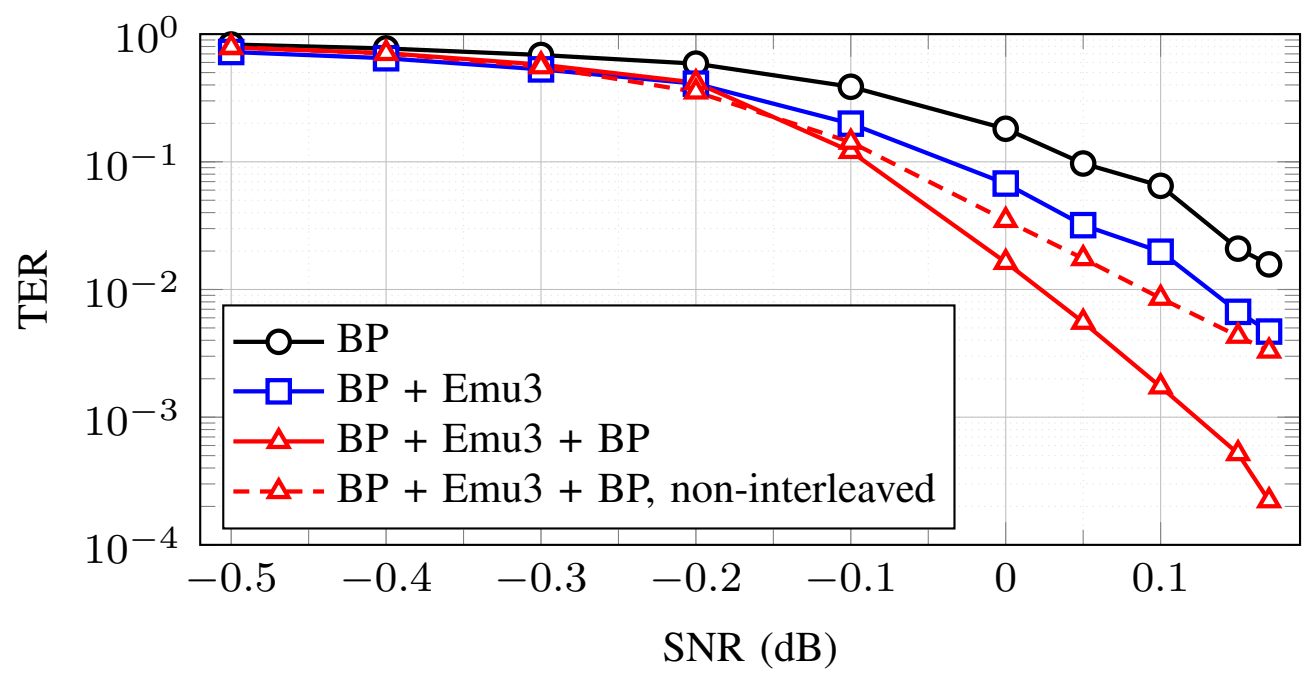
\begin{figure}[!t]
\centering
\input{figures/tikz/fig_ter}
\vspace{-1em}
\caption{TER \blue{on Kodak24 after the first BP pass (BP), token recovery (BP + Emu3) and the second BP pass (BP + Emu3 + BP)} at $\cpp=0.5$.}
\vspace{-0.8em}
\label{fig:interleaving-fixed}
\end{figure}

\begin{figure}[!t]
\centering
\input{figures/tikz/fig_psnr}
\vspace{-1em}
\caption{PSNR \blue{on Kodak24} at $\cpp=0.5$. \blue{The proposed receiver is BP + Emu3 + BP with interleaving.}}
\vspace{-0.8em}
\label{fig:snr-results}
\end{figure}
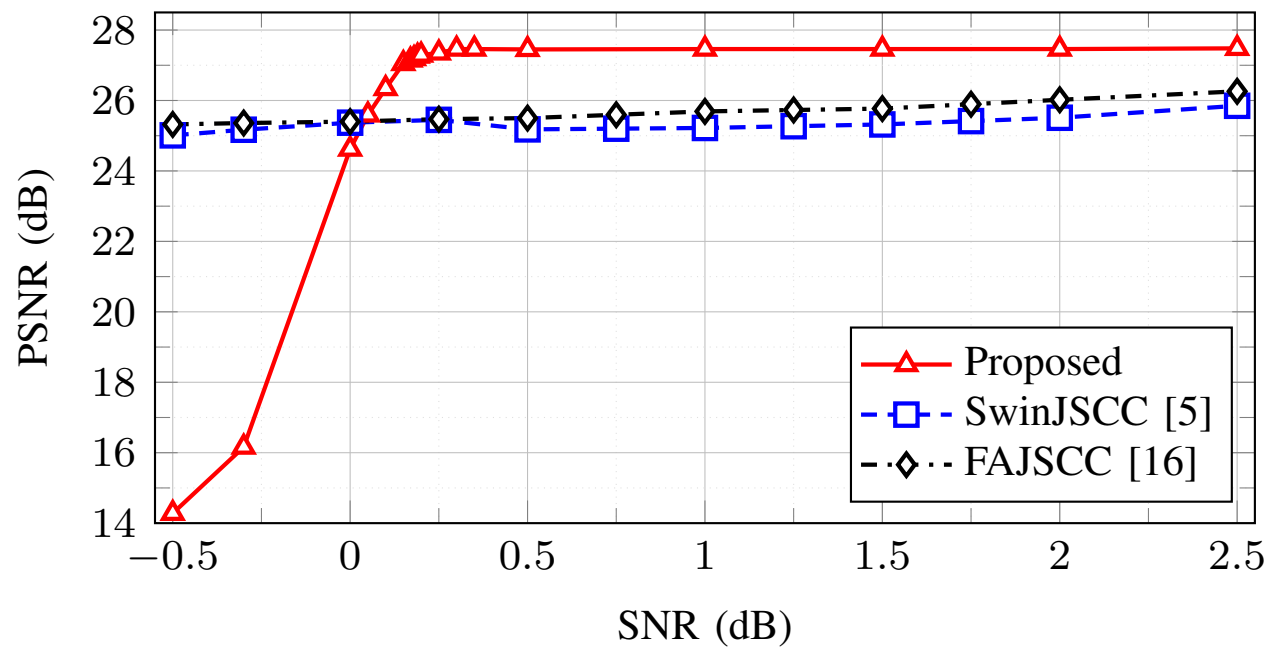

\begin{figure*}[!t]
\centering
\input{figures/tikz/fig_visual}
\vspace{-1em}
\caption{ \blue{Reconstructions of Kodak16 at an SNR of 0.1~dB and $\cpp=0.5$, with their PSNR.}}
\vspace{-1em}
\label{fig:reconstruction}
\end{figure*}

\vspace{-0.8em}
\section{Experimental Results}\label{sec:experiments}

\vspace{-0.3em}

\subsection{\protect\blue{Experimental Setup}}\label{ssec:setup}
\vspace{-0.3em}

\blue{We evaluate on Kodak24, with the images resized to $512\times512$ pixels. The Emu3 tokenizer represents each image by $T=4096$ indices on a $64\times64$ grid with $V=2^{14}$, i.e., $M=14$, and the 57,344 information bits are partitioned into $B=7$ blocks of $K=8192$ bits. Each block is encoded by the 5G NR LDPC code with base graph 1 in Sionna~\cite{hoydis2022sionna} with a coded length $N$ such that $\cpp=0.5$, i.e., $R\approx0.44$, and each BP pass runs 20 iterations. Emu3-Gen is fine-tuned on 25,600 pairs of hard and clean token sequences from 5,929 crops of the DIV2K training set and Flickr2K, for 30 epochs of 3,200 pairs each, and the detokenizer on 4,480 clean token--image pairs. Both models are validated on the DIV2K validation set, and no Kodak image is used in training. We set $\lambda=1$. All experiments run on an NVIDIA RTX 5090 GPU.}

\blue{The receivers are named after their stages. BP reconstructs the image from $\tilde{\mathbf s}$, BP + Emu3 from $\hat{\mathbf s}$, and the proposed receiver BP + Emu3 + BP from $\hat{\mathbf s}^{\mathrm{BP}}$. All three use interleaving and the fine-tuned detokenizer, and the non-interleaved variant of the proposed receiver uses the identity permutation and the same Emu3-Gen. SwinJSCC~\cite{yang2023swinjscc} and FAJSCC~\cite{choi2025fajscc} are fine-tuned at $\cpp=0.5$ as in~\cite{choi2025fajscc} and evaluated at the same SNR per real channel dimension. We report the PSNR, the LPIPS with AlexNet features~\cite{zhang2018lpips} and the TER, i.e., the fraction of erroneous token indices.}

\vspace{-0.3em}
\subsection{\protect\blue{Two-Pass BP Schedule and Interleaving}}\label{ssec:ablation}
\vspace{-0.3em}

\blue{Fig.~\ref{fig:interleaving-fixed} shows the TER after each stage of the receiver. At 0.1~dB, token recovery reduces the TER from 0.065 to 0.020, and the second pass reduces it to $1.7\times10^{-3}$. At 0.15~dB, the corresponding values are 0.021, $6.7\times10^{-3}$ and $5.2\times10^{-4}$. Token recovery thus removes 70\% and 68\% of the token errors, and the second pass removes 91\% and 92\% of the remaining ones, since the prior LLRs of a failed block are wrong only at these remaining errors, which BP corrects with the parity constraints. At $-0.2$~dB and below, more than 40\% of the tokens remain erroneous after token recovery, and the second pass increases the TER, e.g., from 0.53 to 0.58 at $-0.3$~dB.}

\blue{Interleaving reduces the final TER from $8.6\times10^{-3}$ to $1.7\times10^{-3}$ at 0.1~dB and from $4.3\times10^{-3}$ to $5.2\times10^{-4}$ at 0.15~dB, i.e., by 80\% and 88\%, although the first pass leaves the same expected number of erroneous tokens for both orders (Section~\ref{ssec:bp}). By \eqref{eq:indep-tokens}, the neighbors of an erroneous token are then mostly correct, and token recovery predicts the token from them. At $-0.2$~dB and below, where most blocks fail, the raster order gives a lower final TER, e.g., 0.35 versus 0.42 at $-0.2$~dB.}

\subsection{Comparison with JSCC Baselines}\label{ssec:baseline-results}

\begin{table}[!t]
\centering
\caption{\blue{Average PSNR and LPIPS on Kodak24 at $\cpp=0.5$ over the SNRs from 0 to 2.5~dB.}}
\label{tab:psnr-avg}
\setlength{\tabcolsep}{4pt}
\begin{tabular}{lcc}
\toprule
\blue{Receiver} & \blue{PSNR (dB) $\uparrow$} & \blue{LPIPS $\downarrow$}\\\midrule
\blue{SwinJSCC~\cite{yang2023swinjscc}} & \blue{25.38} & \blue{0.36}\\
\blue{FAJSCC~\cite{choi2025fajscc}} & \blue{25.73} & \blue{0.43}\\
\blue{BP} & \blue{25.24} & \blue{0.13}\\
\blue{BP + Emu3 + BP, non-interleaved} & \blue{27.03} & \blue{0.093}\\
\blue{BP + Emu3 + BP (proposed)} & \blue{\textbf{27.17}} & \blue{\textbf{0.067}}\\\bottomrule
\end{tabular}
\vspace{-0.8em}
\end{table}

\blue{In Fig.~\ref{fig:snr-results}, the PSNR of the proposed receiver rises from 14.3~dB at $-0.5$~dB, where BP fails on most blocks, to 27.5~dB at 0.3~dB, and stays at 27.5~dB, the PSNR with error-free tokens, up to 2.5~dB. The PSNR of the JSCC baselines lies between 25.0 and 26.3~dB over the whole range. From 0.25~dB to 2.5~dB, i.e., at medium to high SNR, the proposed receiver exceeds SwinJSCC by at least 1.63~dB and FAJSCC by at least 1.22~dB. Table~\ref{tab:psnr-avg} averages the PSNR and the LPIPS over the SNRs from 0 to 2.5~dB at which the baselines are evaluated. The proposed receiver attains the highest PSNR, 1.79~dB above SwinJSCC and 1.44~dB above FAJSCC, and the lowest LPIPS, 0.067, whereas the blurred JSCC reconstructions have LPIPS of 0.36 and 0.43. Interleaving increases the PSNR by 0.14~dB and reduces the LPIPS from 0.093 to 0.067.}

\blue{In Fig.~\ref{fig:reconstruction}, at 0.1~dB, BP leaves scattered artifacts, and the JSCC reconstructions blur the clouds and the vegetation. The proposed receiver removes the artifacts, keeps the clouds and the palm trees sharp, and exceeds the PSNR of SwinJSCC by 0.99~dB and that of FAJSCC by 0.87~dB.}

\vspace{-0.5em}
\section{Conclusion}
\vspace{-0.5em}

\blue{We proposed spatial token interleaving and a two-pass BP schedule for images transmitted as VQ tokens in multiple 5G NR LDPC blocks. The interleaver makes the errors of adjacent tokens independent without additional transmitted bits, and the second BP pass decodes each block again with the fused token distributions as priors. On Kodak24 at $\cpp=0.5$ and 0.15~dB, the second pass removes 92\% of the token errors left by token recovery, and interleaving reduces the final TER by 88\%. At medium to high SNR, the proposed receiver exceeds both JSCC baselines by at least 1.2~dB in PSNR.}

\bibliographystyle{IEEEtran}
\bibliography{ref}
\end{document}

%% file: figures/tikz/fig_system_2.tex
\begingroup
\definecolor{figInk}{RGB}{38,49,62}
\definecolor{figBlue}{RGB}{51,103,153}
\definecolor{figTeal}{RGB}{32,125,120}
\definecolor{figAmber}{RGB}{173,122,47}
\begin{tikzpicture}[
  x=1cm,y=1cm,
  font=\fontsize{8}{9.2}\selectfont,text=figInk,
  line join=round,line cap=round,
  >={Stealth[length=3pt,width=2.3pt]},
  block/.style={draw=figBlue,line width=.55pt,fill=white,
    rounded corners=1.3pt,align=center,inner sep=2pt,
    minimum height=.90cm},
  flow/.style={->,draw=figInk,line width=.55pt},
  softflow/.style={flow,draw=figBlue},
  signal/.style={font=\fontsize{7.5}{8.5}\selectfont,
    inner sep=1pt},
  symbolcell/.style={rectangle,draw=figBlue!80,fill=figBlue!12,
    line width=.35pt,minimum size=.14cm,inner sep=0pt,outer sep=0pt},
  groupframe/.style={line width=.45pt,rounded corners=3pt},
  title/.style={font=\bfseries\fontsize{8.5}{10}\selectfont,
    anchor=west},
  imagecaption/.style={font=\fontsize{7.5}{8.5}\selectfont,
    align=center,inner sep=0pt}
]
\path[use as bounding box] (-.14,-3.74) rectangle (8.58,1.00);
\def\TxY{0}
\def\RxY{-2.60}
\draw[groupframe,draw=figBlue!30,fill=figBlue!3]
  (-.12,-1.06) rectangle (7.12,.98);
\draw[groupframe,draw=figTeal!30,fill=figTeal!3]
  (-.12,-3.72) rectangle (7.12,-1.56);
\node[title,text=figBlue] at (0,.72) {Transmitter};
\node[title,text=figTeal] at (0,-1.82) {Receiver};

\node[inner sep=0pt] (source) at (.42,\TxY)
  {\includegraphics[width=.72cm,height=.72cm]{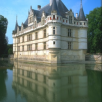}};
\draw[figInk!35,line width=.3pt]
  (source.south west) rectangle (source.north east);
\node[imagecaption] at (.42,-.68) {Input\\$\mathbf X$};
\node[block,fill=figBlue!4,minimum width=1.18cm]
  (vqencoder) at (1.78,\TxY) {Visual\\encoder};
\node[block,minimum width=1.13cm]
  (ldpcencoder) at (3.90,\TxY) {LDPC\\encoding};
\node[block,minimum width=1.50cm]
  (modulator) at (6.05,\TxY) {BPSK\\modulation};
\draw[flow] (source.east) -- (vqencoder.west);
\draw[flow] (vqencoder.east) --
  node[signal,above] {$\mathbf u_j$} (ldpcencoder.west);
\draw[flow] (ldpcencoder.east) --
  node[signal,above] {$\mathbf c_j$} (modulator.west);

\node[block,draw=figAmber,fill=figAmber!5,minimum width=1.12cm]
  (channel) at (7.95,-1.30) {AWGN\\channel};
\draw[flow] (modulator.east) -- (7.95,\TxY) -- (channel.north);
\foreach \k in {0,...,3}
  \node[symbolcell] at ({7.26+.18*\k},.14) {};
\node[signal,above] at (7.53,.27) {$x_{j,n}$};

\node[block,minimum width=1.55cm]
  (demodulator) at (6.05,\RxY) {Soft\\demodulation};
\node[block,minimum width=1.13cm]
  (bpdecoder) at (3.90,\RxY) {BP\\decoding};
\node[block,draw=figTeal,fill=figTeal!4,minimum width=1.18cm]
  (vqdecoder) at (1.78,\RxY) {Visual\\decoder};
\draw[flow] (channel.south) -- (7.95,\RxY) -- (demodulator.east);
\foreach \k in {0,...,3}
  \node[symbolcell,draw=figTeal!80,fill=figTeal!12]
    at ({7.26+.18*\k},{\RxY+.14}) {};
\node[signal,above] at (7.53,{\RxY+.27}) {$y_{j,n}$};
\draw[softflow] (demodulator.west) --
  node[signal,above] {$\ell^{\mathrm{ch}}_{j,n}$} (bpdecoder.east);
\draw[softflow] (bpdecoder.west) --
  node[signal,above] {$\boldsymbol\ell_j^{\mathrm{APP}}$} (vqdecoder.east);

\node[inner sep=0pt] (reconstruction) at (.42,\RxY)
  {\includegraphics[width=.72cm,height=.72cm]{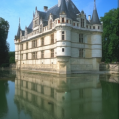}};
\draw[figInk!35,line width=.3pt]
  (reconstruction.south west) rectangle (reconstruction.north east);
\node[imagecaption] at (.42,-3.28) {Output\\$\hat{\mathbf X}$};
\draw[flow] (vqdecoder.west) -- (reconstruction.east);
\draw[softflow] (demodulator.south) -- (6.05,-3.42) -- (1.78,-3.42)
  -- (vqdecoder.south);
\node[signal,below] at (3.90,-3.42) {$\ell^{\mathrm{ch}}_{j,n}$ (second BP pass)};
\end{tikzpicture}
\endgroup

%% file: figures/tikz/fig_encoder_2.tex
\begingroup%
\definecolor{encInk}{RGB}{38,49,62}%
\definecolor{encBlue}{RGB}{51,103,153}%
\definecolor{encTeal}{RGB}{32,125,120}%
\definecolor{encAmber}{RGB}{179,119,38}%
\definecolor{encPurple}{RGB}{117,93,157}%
\definecolor{encGray}{RGB}{231,235,240}%
\begin{tikzpicture}[
  x=1cm,y=1cm,font=\fontsize{8}{9.2}\selectfont,text=encInk,
  line cap=round,line join=round,
  >={Stealth[length=3pt,width=2.3pt]},
  flow/.style={->,draw=encInk,line width=.55pt},
  signal/.style={font=\fontsize{7.7}{8.8}\selectfont,inner sep=1pt},
  bitvalue/.style={font=\fontsize{6.7}{7.2}\selectfont,inner sep=0pt},
  bitindex/.style={font=\fontsize{6.7}{7.2}\selectfont,inner sep=0pt},
  bitboundary/.style={draw=encBlue,line width=.65pt,
    dash pattern=on 1.6pt off 1.1pt}
]
\path[use as bounding box] (-.02,-4.50) rectangle (8.78,1.25);
\draw[draw=encBlue!30,fill=encBlue!2,line width=.45pt,
  rounded corners=2pt] (0,-3.47) rectangle (8.76,1.23);
\node[anchor=west,text=encBlue,inner sep=0pt,
  font=\bfseries\fontsize{8.5}{9.5}\selectfont] at (.18,1.00) {Visual encoder $\E$};

\node[inner sep=0pt] (source) at (.88,0)
  {\includegraphics[width=1.12cm,height=1.12cm]{figures/tikz/assets/source.png}};
\node[signal] at (.99,-.77) {$\mathbf X\in\mathbb R^{H\times W\times3}$};
\node[inner sep=0pt] (spatial) at (2.46,0)
  {\includegraphics[width=1.12cm,height=1.12cm]{figures/tikz/assets/source.png}};
\foreach \x in {2.27333,2.64667}{
  \draw[encInk!75,line width=1pt] (\x,-.56) -- (\x,.56);
  \draw[white,line width=.55pt] (\x,-.56) -- (\x,.56);
}
\foreach \y in {-.18667,.18667}{
  \draw[encInk!75,line width=1pt] (1.90,\y) -- (3.02,\y);
  \draw[white,line width=.55pt] (1.90,\y) -- (3.02,\y);
}
\draw[encInk!35,line width=.3pt]
  (source.south west) rectangle (source.north east);
\draw[encInk!35,line width=.3pt]
  (spatial.south west) rectangle (spatial.north east);
\draw[flow] (source.east) -- (spatial.west);

\node[draw=encPurple,fill=encPurple!6,line width=.6pt,
  rounded corners=1.5pt,minimum width=1.70cm,minimum height=1.12cm,
  inner sep=2pt,align=center]
  (vq) at (4.20,0) {Emu3 VQ\\tokenizer\\$\mathcal{E}_{\mathrm{VQ}}$};
\draw[flow] (spatial.east) -- (vq.west);

\def\gridLeft{5.64}
\def\gridTop{.54}
\def\gridWidth{.66}
\def\gridHeight{.36}
\foreach \col/\row/\entry in {
  0/0/s_1,1/0/s_2,2/0/s_3,3/0/s_4,
  0/1/\vdots,1/1/\vdots,2/1/\ddots,3/1/\vdots,
  0/2/s_t,1/2/\cdots,2/2/\cdots,3/2/s_T}{
  \def\tokenFill{encGray}
  \def\tokenEdge{encInk!42}
  \ifnum\row=0
    \ifnum\col=0 \def\tokenFill{encTeal!24}\def\tokenEdge{encTeal}\fi
    \ifnum\col=1 \def\tokenFill{encAmber!27}\def\tokenEdge{encAmber}\fi
    \ifnum\col=2 \def\tokenFill{encPurple!23}\def\tokenEdge{encPurple}\fi
  \fi
  \path[draw=\tokenEdge,fill=\tokenFill,line width=.4pt]
    ({\gridLeft+\col*\gridWidth},{\gridTop-\row*\gridHeight})
    rectangle ++(\gridWidth,-\gridHeight);
  \node[signal] at
    ({\gridLeft+(\col+.5)*\gridWidth},
     {\gridTop-(\row+.5)*\gridHeight}) {$\entry$};
}
\node[signal] at (6.96,.77) {Token indices};
\draw[flow] (vq.east) -- (5.64,0);

\node[draw=encTeal,fill=encTeal!6,line width=.55pt,
  rounded corners=1.5pt,minimum width=3.80cm,minimum height=.47cm,
  inner sep=2pt] (interleaver) at (4.38,-1.05) {Token interleaver};
\draw[flow] (6.96,-.54) -- (6.96,-.68)
  -- (4.38,-.68) -- (interleaver.north);

\def\streamLeft{.46}
\def\tokenWidth{.98}
\foreach \entry/\shade [count=\slot from 0] in {
  s_t/0,s_1/1,s_T/0,\cdots/0,s_4/0,s_2/2,\cdots/0,s_3/3}{
  \def\tokenFill{encGray}
  \def\tokenEdge{encInk!42}
  \ifnum\shade=1 \def\tokenFill{encTeal!24}\def\tokenEdge{encTeal}\fi
  \ifnum\shade=2 \def\tokenFill{encAmber!27}\def\tokenEdge{encAmber}\fi
  \ifnum\shade=3 \def\tokenFill{encPurple!23}\def\tokenEdge{encPurple}\fi
  \path[draw=\tokenEdge,fill=\tokenFill,line width=.45pt]
    ({\streamLeft+\slot*\tokenWidth},-1.52) rectangle ++(\tokenWidth,-.42);
  \node at ({\streamLeft+(\slot+.5)*\tokenWidth},-1.73) {$\entry$};
}
\draw[flow] (interleaver.south) -- (4.38,-1.52);
\draw[flow] (4.38,-1.94) -- (4.38,-2.40);
\node[signal,anchor=west] at (4.45,-2.17) {Binary mapping $\mathbf q(\cdot)$};

\begin{scope}[yshift=.17cm]
\pgfmathsetmacro{\bitWidth}{\tokenWidth/4}
\pgfmathsetmacro{\blockWidth}{6*\bitWidth}
\pgfmathsetmacro{\startOne}{\streamLeft}
\pgfmathsetmacro{\startTwo}{\streamLeft+\blockWidth}
\pgfmathsetmacro{\startMiddle}{\streamLeft+6*\tokenWidth-\blockWidth}
\pgfmathsetmacro{\startLast}{\streamLeft+8*\tokenWidth-\blockWidth}
\pgfmathsetmacro{\gapOne}{(\startTwo+\blockWidth+\startMiddle)/2}
\pgfmathsetmacro{\gapTwo}{(\startMiddle+\blockWidth+\startLast)/2}
\def\bitTop{-2.89}
\def\bitBottom{-3.18}
\foreach \value [count=\bit from 0] in {0,1,1,0,1,0,1,1,1,0,0,0}{
  \def\bitFill{encGray}
  \def\bitEdge{encInk!42}
  \ifnum\bit>3
    \ifnum\bit<8 \def\bitFill{encTeal!24}\def\bitEdge{encTeal}\fi
  \fi
  \path[draw=\bitEdge,fill=\bitFill,line width=.35pt]
    ({\streamLeft+\bit*\bitWidth},\bitTop) rectangle ++(\bitWidth,\bitBottom-\bitTop);
  \node[bitvalue] at
    ({\streamLeft+(\bit+.5)*\bitWidth},{(\bitTop+\bitBottom)/2}) {$\value$};
}
\foreach \start/\kind/\values in {
  \startMiddle/1/{1,0,0,1,1,0},\startLast/2/{0,1,1,1,0,0}}{
  \foreach \value [count=\bit from 0] in \values{
    \def\bitFill{encGray}
    \def\bitEdge{encInk!42}
    \ifnum\kind=1
      \ifnum\bit>1 \def\bitFill{encAmber!27}\def\bitEdge{encAmber}\fi
    \fi
    \ifnum\kind=2
      \ifnum\bit>1 \def\bitFill{encPurple!23}\def\bitEdge{encPurple}\fi
    \fi
    \path[draw=\bitEdge,fill=\bitFill,line width=.35pt]
      ({\start+\bit*\bitWidth},\bitTop)
      rectangle ++(\bitWidth,\bitBottom-\bitTop);
    \node[bitvalue] at
      ({\start+(\bit+.5)*\bitWidth},{(\bitTop+\bitBottom)/2}) {$\value$};
  }
}
\node[signal] at ({\streamLeft+1.5*\tokenWidth},-2.69) {$\mathbf q(s_1)$};
\node[signal] at ({\streamLeft+5.5*\tokenWidth},-2.69) {$\mathbf q(s_2)$};
\node[signal] at ({\streamLeft+7.5*\tokenWidth},-2.69) {$\mathbf q(s_3)$};
\foreach \edge in {7,8}{
  \draw[encBlue,line width=.65pt]
    ({\streamLeft+\edge*\tokenWidth},\bitTop)
    -- ++(0,.06);
}
\node[signal] at (\gapOne,-3.025) {$\cdots$};
\node[signal] at (\gapTwo,-3.025) {$\cdots$};
\foreach \start/\label in {\startOne/1,\startTwo/2,\startMiddle/j,\startLast/B}{
  \draw[bitboundary] (\start,-2.84) -- (\start,-3.23);
  \draw[bitboundary] ({\start+\blockWidth},-2.84)
    -- ({\start+\blockWidth},-3.23);
  \draw[encBlue,line width=.5pt]
    ({\start+.035},-3.25) -- ++(0,-.07)
    -- ++(\blockWidth-.07,0) -- ++(0,.07);
  \node[signal,text=encBlue] at ({\start+\blockWidth/2},-3.46)
    {$\mathbf u_{\label}$};
}
\node[signal,text=encBlue] at (\gapOne,-3.46) {$\cdots$};
\node[signal,text=encBlue] at (\gapTwo,-3.46) {$\cdots$};
\foreach \start/\label in {\startOne/1,\startTwo/2,\startMiddle/j,\startLast/B}{
  \draw[flow] ({\start+\blockWidth/2},-3.57) -- ({\start+\blockWidth/2},-3.73);
  \node[draw=encBlue,fill=encBlue!6,line width=.5pt,rounded corners=1.5pt,
    minimum width={\blockWidth*1cm-.12cm},minimum height=.50cm,inner sep=1pt,
    align=center,font=\fontsize{7.2}{7.9}\selectfont]
    at ({\start+\blockWidth/2},-3.98) {5G LDPC\\encoder};
  \draw[flow] ({\start+\blockWidth/2},-4.23) -- ({\start+\blockWidth/2},-4.40);
  \node[signal,anchor=north] at ({\start+\blockWidth/2},-4.38) {$\mathbf c_{\label}$};
}
\node[signal] at (\gapOne,-3.98) {$\cdots$};
\node[signal] at (\gapTwo,-3.98) {$\cdots$};
\end{scope}
\end{tikzpicture}%
\endgroup%

%% file: figures/tikz/fig_decoder_2.tex
\begingroup%
\definecolor{decInk}{RGB}{38,49,62}%
\definecolor{decBlue}{RGB}{51,103,153}%
\definecolor{decTeal}{RGB}{32,125,120}%
\definecolor{decPurple}{RGB}{117,93,157}%
\definecolor{decAmber}{RGB}{179,119,38}%
\definecolor{decGray}{RGB}{231,235,240}%
\begin{tikzpicture}[
  x=1cm,y=1cm,font=\fontsize{8}{9.2}\selectfont,text=decInk,
  line cap=round,line join=round,
  >={Stealth[length=3pt,width=2.3pt]},
  block/.style={draw=decTeal,line width=.55pt,fill=white,
    rounded corners=1.5pt,align=center,inner sep=2pt},
  stage/.style={draw=decTeal!65,fill=decBlue!6,line width=.6pt,
    rounded corners=2pt},
  stageheading/.style={anchor=east,text=decTeal,fill=white,
    font=\bfseries\fontsize{8}{9.2}\selectfont,
    inner xsep=4pt,inner ysep=2pt},
  flow/.style={->,draw=decInk,line width=.55pt},
  softflow/.style={flow,draw=decBlue},
  arflow/.style={flow,draw=decPurple},
  feedback/.style={arflow,dash pattern=on 2pt off 1.4pt},
  signal/.style={font=\fontsize{7.4}{8.5}\selectfont,inner sep=1pt},
  smallnote/.style={font=\fontsize{7.2}{8.3}\selectfont,inner sep=1pt},
  llrboundary/.style={draw=decBlue,line width=.65pt,
    dash pattern=on 1.6pt off 1.1pt},
  heading/.style={anchor=west,text=decTeal,
    font=\bfseries\fontsize{8}{9.2}\selectfont,inner sep=0pt}
]
\path[use as bounding box] (-.02,-8.09) rectangle (8.78,2.42);
\node[heading] at (.18,2.26) {First BP pass};
\foreach \cx/\label in {1.075/1,2.745/2,5.395/j,7.705/B}{
  \draw[softflow] (\cx,2.10) -- (\cx,1.83);
  \node[signal,anchor=west,text=decBlue] at ({\cx+.05},1.97)
    {$\boldsymbol\ell^{\mathrm{ch}}_{\label}$};
  \node[block,draw=decBlue,fill=decBlue!13,minimum width=1.55cm,
    minimum height=.40cm,inner sep=1pt,font=\fontsize{7.4}{8.2}\selectfont]
    at (\cx,1.62) {BP decoding};
  \draw[softflow,-] (\cx,1.42) -- (\cx,.98);
  \node[signal,anchor=west,text=decBlue] at ({\cx+.05},1.19)
    {$\boldsymbol\ell^{\mathrm{APP}}_{\label}$};
}
\node[signal] at (4.07,1.62) {$\cdots$};
\node[signal] at (6.55,1.62) {$\cdots$};
\draw[softflow,-] (1.075,.98) -- (7.705,.98);
\draw[draw=decTeal!30,fill=decTeal!2,line width=.45pt,
  rounded corners=2pt] (0,-8.07) rectangle (8.76,.80);
\node[heading] at (.18,.58) {Visual decoder $\mathcal D$};
\node[block,minimum width=4.10cm,minimum height=.47cm]
  (deinterleave) at (4.38,.25) {Regroup into tokens, deinterleave};
\draw[softflow] (4.38,.98) -- (deinterleave.north);
\coordinate (spatialLLRs) at (deinterleave.south);

\begin{scope}[yshift=.35cm]
\draw[stage] (.18,-4.40) rectangle (8.60,-.62);
\node[stageheading,anchor=west] at (.32,-.62) {Token recovery};

\coordinate (observations) at (4.38,-.89);
\draw[softflow,-] (spatialLLRs) -- (observations)
  node[pos=.36,signal,text=decBlue,fill=white,
    inner xsep=3pt,inner ysep=1pt]
  {$\mathbf L\in\mathbb R^{T\times M}$};
\fill[decBlue] (observations) circle (1.1pt);

\node[block,draw=decBlue,fill=decBlue!13,
  minimum width=3.14cm,minimum height=.83cm]
  (bp) at (1.94,-2.45)
  {\textcolor{decBlue}{\bfseries BP token scores}\\[2pt]
   $\log\tilde p_{\mathrm{BP}}(v\mid\boldsymbol\ell_t)$};
\draw[softflow] (observations) -- (1.94,-.89) -- (bp.north);

\node[block,draw=decPurple,fill=decPurple!3,minimum width=3.10cm,
  minimum height=.65cm,font=\fontsize{7.4}{8.6}\selectfont]
  (hard) at (5.81,-1.42)
  {Hard bit decisions\\
   $\tilde b_{t,m}=\mathbf 1\{L_{t,m}>0\}$};
\draw[softflow] (observations) -- (7.82,-.89)
  -- (7.82,-1.42) -- (hard.east);
\node[block,draw=decPurple,fill=decPurple!16,
  minimum width=3.10cm,minimum height=.83cm]
  (ar) at (5.81,-2.45)
  {\textcolor{decPurple}{\bfseries Emu3-Gen (AR)}\\[2pt]
   $p_\theta(v\mid\hat{\mathbf s}_{<t},\tilde{\mathbf s})$};
\draw[arflow] (hard.south) -- (ar.north)
  node[midway,signal,right,text=decPurple] {$\tilde{\mathbf s}$};

\node[block,draw=decTeal,fill=decTeal!13,
  minimum width=6.88cm,minimum height=.98cm,inner sep=3pt]
  (fusion) at (3.98,-3.66)
  {{\bfseries Token fusion: $t=1,\ldots,T$ in raster order}\\[3pt]
   {\fontsize{7.5}{8.5}\selectfont
   $\displaystyle p_t^{\mathrm F}(v)=\operatorname{softmax}_v
      \bigl\{\log\tilde p_{\mathrm{BP}}(v\mid\boldsymbol\ell_t)
      +\lambda\log p_\theta(v\mid\hat{\mathbf s}_{<t},\tilde{\mathbf s})\bigr\}$}};
\draw[softflow] (bp.south) -- (bp.south |- fusion.north);
\draw[arflow] (ar.south) -- (ar.south |- fusion.north);
\draw[feedback] (fusion.east) -- (8.48,-3.66)
  -- (8.48,-2.45) -- (ar.east);
\node[smallnote,align=center,text=decPurple] at (8.04,-3.07)
  {Append\\$\hat s_t$\\next $t$};

\draw[stage] (.18,-6.42) rectangle (8.60,-4.70);
\node[stageheading,anchor=west] at (.32,-4.70) {Second BP pass};
\node[block,minimum width=2.38cm,minimum height=.80cm,font=\fontsize{7}{8}\selectfont]
  (bitPrior) at (1.65,-5.55)
  {Bit marginalization\\$+$ LLR conversion};
\node[block,minimum width=2.24cm,minimum height=.80cm,font=\fontsize{7}{8}\selectfont]
  (reorder) at (4.36,-5.55)
  {Interleave and\\regroup into blocks};
\node[block,draw=decBlue,fill=decBlue!13,
  minimum width=2.24cm,minimum height=.80cm,font=\fontsize{7}{8}\selectfont]
  (bpAgain) at (7.10,-5.55)
  {BP decoding};
\draw[softflow] (fusion.south) -- (3.98,-4.96)
  -- (1.65,-4.96) -- (bitPrior.north);
\draw[softflow] (bitPrior.east) -- (reorder.west);
\draw[softflow] (reorder.east) -- (bpAgain.west);
\draw[softflow] (7.10,-4.84) -- (bpAgain.north)
  node[pos=.45,signal,right,text=decBlue] {$\boldsymbol\ell^{\mathrm{ch}}$};
\node[smallnote,text=decBlue,font=\fontsize{6.7}{7.7}\selectfont] at (4.20,-6.22)
  {Prior LLRs for information bits, channel LLRs for parity bits};

\def\recoveredGridLeft{.64}
\def\recoveredGridTop{-6.98}
\def\recoveredCellWidth{.66}
\def\recoveredCellHeight{.36}
\node[inner sep=0pt,minimum width=2.64cm,minimum height=1.08cm]
  (recoveredGrid) at (1.96,-7.52) {};
\foreach \col/\row/\entry in {
  0/0/\hat s^{\mathrm{BP}}_1,1/0/\hat s^{\mathrm{BP}}_2,2/0/\hat s^{\mathrm{BP}}_3,3/0/\hat s^{\mathrm{BP}}_4,
  0/1/\vdots,1/1/\vdots,2/1/\ddots,3/1/\vdots,
  0/2/\hat s^{\mathrm{BP}}_t,1/2/\cdots,2/2/\cdots,3/2/\hat s^{\mathrm{BP}}_T}{
  \def\recoveredFill{decGray}
  \def\recoveredEdge{decInk!42}
  \ifnum\row=0
    \ifnum\col=0 \def\recoveredFill{decTeal!24}\def\recoveredEdge{decTeal}\fi
    \ifnum\col=1 \def\recoveredFill{decAmber!27}\def\recoveredEdge{decAmber}\fi
    \ifnum\col=2 \def\recoveredFill{decPurple!23}\def\recoveredEdge{decPurple}\fi
  \fi
  \path[draw=\recoveredEdge,fill=\recoveredFill,line width=.4pt]
    ({\recoveredGridLeft+\col*\recoveredCellWidth},
     {\recoveredGridTop-\row*\recoveredCellHeight})
    rectangle ++(\recoveredCellWidth,-\recoveredCellHeight);
  \node[signal] at
    ({\recoveredGridLeft+(\col+.5)*\recoveredCellWidth},
     {\recoveredGridTop-(\row+.5)*\recoveredCellHeight}) {$\entry$};
}
\node[block,draw=decTeal,fill=decTeal!16,
  minimum width=2.00cm,minimum height=1.08cm]
  (imageDecoder) at (5.79,-7.52)
  {Emu3 VQ\\detokenizer\\$\mathcal D_{\mathrm{VQ}}$};
\draw[flow] (bpAgain.south) -- (7.10,-6.04) -- (8.42,-6.04) -- (8.42,-6.72)
  -- (1.96,-6.72) -- (recoveredGrid.north);
\node[signal,fill=decTeal!2,inner sep=2pt] at (4.38,-6.72)
  {Hard decisions $\to$ restore spatial token order};
\draw[flow] (recoveredGrid.east) -- (imageDecoder.west);
\node[inner sep=0pt] (output) at (7.94,-7.52)
  {\includegraphics[width=.86cm,height=.86cm]{figures/tikz/assets/reconstruction.png}};
\draw[decInk!35,line width=.3pt]
  (output.south west) rectangle (output.north east);
\draw[flow] (imageDecoder.east) -- (output.west);
\node[signal] at (7.94,-8.14) {Output $\hat{\mathbf X}$};
\end{scope}
\end{tikzpicture}%
\endgroup%

%% file: figures/tikz/fig_ter.tex
\begin{tikzpicture}
\begin{semilogyaxis}[
  width=\columnwidth, height=0.56\columnwidth,
  xmin=-0.52, xmax=0.19, ymin=1e-4, ymax=1,
  ytick={1e-4,1e-3,1e-2,1e-1,1},
  xtick={-0.5,-0.4,-0.3,-0.2,-0.1,0,0.1},
  xticklabels={$-0.5$,$-0.4$,$-0.3$,$-0.2$,$-0.1$,$0$,$0.1$},
  minor x tick num=1,
  xlabel={SNR (dB)}, ylabel={TER},
  grid=both,
  major grid style={line width=.2pt,draw=gray!50},
  minor grid style={line width=.2pt,draw=gray!25,densely dotted},
  tick label style={font=\footnotesize},
  label style={font=\footnotesize},
  legend style={font=\footnotesize,at={(0.02,0.03)},anchor=south west,
    draw=black,fill=white,inner sep=1.5pt,row sep=-1.5pt},
  legend cell align=left,
  every axis plot/.append style={line width=.8pt,mark size=2.1pt,
    mark options={solid,fill=white}},
]
\addplot[black,mark=*] coordinates {
  (-0.5,0.8296) (-0.4,0.7725) (-0.3,0.6864) (-0.2,0.5871) (-0.1,0.3871)
  (0,0.1815) (0.05,0.09699) (0.1,0.06485) (0.15,0.02089) (0.17,0.01573)};
\addlegendentry{BP}
\addplot[blue,mark=square*] coordinates {
  (-0.5,0.7246) (-0.4,0.6473) (-0.3,0.5296) (-0.2,0.4092) (-0.1,0.1991)
  (0,0.06761) (0.05,0.03191) (0.1,0.01977) (0.15,0.006683) (0.17,0.004683)};
\addlegendentry{BP + Emu3}
\addplot[red,mark=triangle*] coordinates {
  (-0.5,0.7846) (-0.4,0.7095) (-0.3,0.5771) (-0.2,0.4171) (-0.1,0.1207)
  (0,0.01632) (0.05,0.005547) (0.1,0.001729) (0.15,0.0005188) (0.17,0.0002218)};
\addlegendentry{BP + Emu3 + BP}
\addplot[red,densely dashed,mark=triangle*] coordinates {
  (-0.5,0.7899) (-0.4,0.7066) (-0.3,0.5561) (-0.2,0.3529) (-0.1,0.1426)
  (0,0.03481) (0.05,0.01747) (0.1,0.008553) (0.15,0.004311) (0.17,0.003299)};
\addlegendentry{BP + Emu3 + BP, non-interleaved}
\end{semilogyaxis}
\end{tikzpicture}%

%% file: figures/tikz/fig_psnr.tex
\begin{tikzpicture}
\begin{axis}[
  width=\columnwidth, height=0.56\columnwidth,
  xmin=-0.55, xmax=2.55, ymin=14, ymax=28.5,
  xtick={-0.5,0,0.5,1,1.5,2,2.5},
  ytick={14,16,18,20,22,24,26,28},
  minor x tick num=1, minor y tick num=1,
  xlabel={SNR (dB)}, ylabel={PSNR (dB)},
  grid=both,
  major grid style={line width=.2pt,draw=gray!50},
  minor grid style={line width=.2pt,draw=gray!25,densely dotted},
  tick label style={font=\footnotesize},
  label style={font=\footnotesize},
  legend style={font=\footnotesize,at={(0.98,0.04)},anchor=south east,
    draw=black,fill=white,inner sep=1.5pt,row sep=-1.5pt},
  legend cell align=left,
  every axis plot/.append style={line width=.8pt,mark size=2.1pt,
    mark options={solid,fill=white}},
]
\addplot[red,mark=triangle*] coordinates {
  (-0.5,14.28) (-0.3,16.16) (0,24.62) (0.05,25.59) (0.1,26.34) (0.15,27.05)
  (0.17,27.15) (0.18,27.19) (0.19,27.26) (0.2,27.29) (0.25,27.35) (0.3,27.46)
  (0.35,27.46) (0.5,27.45) (1,27.46) (1.5,27.46) (2,27.46) (2.5,27.48)};
\addlegendentry{Proposed}
\addplot[blue,densely dashed,mark=square*] coordinates {
  (-0.5,25.01) (-0.3,25.17) (0,25.37) (0.25,25.45) (0.5,25.18) (0.75,25.20)
  (1,25.22) (1.25,25.27) (1.5,25.32) (1.75,25.415) (2,25.51) (2.5,25.85)};
\addlegendentry{SwinJSCC~\cite{yang2023swinjscc}}
\addplot[black,dashdotted,mark=diamond*] coordinates {
  (-0.5,25.32) (-0.3,25.36) (0,25.40) (0.25,25.47) (0.5,25.50) (0.75,25.595)
  (1,25.69) (1.25,25.73) (1.5,25.77) (1.75,25.895) (2,26.02) (2.5,26.26)};
\addlegendentry{FAJSCC~\cite{choi2025fajscc}}
\end{axis}
\end{tikzpicture}%

%% file: figures/tikz/fig_visual.tex
\begin{tikzpicture}[x=0.19\textwidth,font=\footnotesize,inner sep=0pt]
\foreach \k/\file/\name/\val in {
  0/original/Original/{},
  1/bp20/BP/{21.46~dB},
  2/swinjscc/{SwinJSCC~\cite{yang2023swinjscc}}/{27.30~dB},
  3/fajscc/{FAJSCC~\cite{choi2025fajscc}}/{27.42~dB},
  4/proposed/Proposed/{28.29~dB}}{
  \node[anchor=north west] (img\k) at (\k,0)
    {\includegraphics[width=0.18\textwidth]{figures/icc/kodak16/\file.png}};
  \node[anchor=south,text height=1.6ex,text depth=.4ex] at ([yshift=2pt]img\k.north) {\name};
  \node[anchor=north,text height=1.6ex,text depth=.4ex] at ([yshift=-2pt]img\k.south) {\val};
}
\end{tikzpicture}%